\documentclass{emulateapj}
\newcommand\pasa{{PASA}}% 

\newcommand{\expnt}[2]{\ensuremath{#1 \times 10^{#2}}}   % scientific notation
\newcommand{\psr}{PSR~J1816+4510}
\newcommand{\gal}{GALEX~J181635.9+451034}

\newcommand{\galex}{\textit{GALEX}}
\newcommand{\swift}{\textit{Swift}}
\newcommand{\fermi}{\textit{Fermi}}
\newcommand{\fgl}{2FGL~J1816.5+4511}
\newcommand{\til}{\raisebox{-0.7ex}{\ensuremath{\tilde{\;}}}}
\newcommand{\ergscm}{\ensuremath{{\rm erg\,s}^{-1}\,{\rm cm}^{-2}}}
\newcommand{\cmsMeV}{\ensuremath{{\rm cm}^{-2}\,{\rm s}^{-1}\,{\rm MeV}^{-1}}}

\begin{document}

\title{Discovery of the Optical/Ultraviolet/Gamma-ray Counterpart to
  the Eclipsing Millisecond Pulsar J1816+4510}
\author{D.~L.~Kaplan\altaffilmark{1,2}, K.~Stovall\altaffilmark{3,4},
  S.~M.~Ransom\altaffilmark{5}, M.~S.~E.~Roberts\altaffilmark{6,7},
  R.~Kotulla\altaffilmark{1}, A.~M.~Archibald\altaffilmark{8}, C.~M.~Biwer\altaffilmark{1},
  J.~Boyles\altaffilmark{9}, L.~Dartez\altaffilmark{3}, D.~F.~Day\altaffilmark{1},
  A.~J.~Ford\altaffilmark{3}, A.~Garcia\altaffilmark{3},
  J.~W.~T.~Hessels\altaffilmark{10,11}, F.~A.~Jenet\altaffilmark{3}, C.~Karako\altaffilmark{8},
  V.~M.~Kaspi\altaffilmark{8}, V.~I.~Kondratiev\altaffilmark{10,12},
  D.~R.~Lorimer\altaffilmark{9,13}, R.~S. Lynch\altaffilmark{8},
  M.~A.~McLaughlin\altaffilmark{9,13}, M.~D.~W.~Rohr\altaffilmark{1}, X.~Siemens\altaffilmark{1},
  I.~H.~Stairs\altaffilmark{14}, \& J.~van~Leeuwen\altaffilmark{10,11}
}

\slugcomment{ApJ, in press}

\altaffiltext{1}{Physics Department, University of
  Wisconsin-Milwaukee, Milwaukee WI 53211, USA; kaplan@uwm.edu}
\altaffiltext{2}{Department of Astronomy, University of
  Wisconsin-Madison, Madison, WI, USA} 
\altaffiltext{3}{Center for Advanced Radio Astronomy and Department of
  Physics and Astronomy, University of Texas at Brownsville,
  Brownsville, TX 78520, USA}
\altaffiltext{4}{Department of Physics and Astronomy, University of
  Texas at San Antonio, San Antonio, TX 78249, USA}
\altaffiltext{5}{National Radio Astronomy Observatory,
  520 Edgemont Road, Charlottesville, VA 22901, US}
\altaffiltext{6}{Eureka Scientific, Inc., 2452 Delmer Street, Suite
  100, Oakland, CA 94602-3017, USA} 
\altaffiltext{7}{Department of Physics, Ithaca College, Ithaca, NY 14850, USA}
\altaffiltext{8}{Department of
  Physics, McGill University, 3600 University Street, Montreal, QC H3A
  2T8, Canada} 
\altaffiltext{9}{Department of Physics, West Virginia
  University, White Hall, 115 Willey Street, Morgantown, WV 26506, USA
} 
\altaffiltext{10}{ASTRON, the Netherlands Institute for Radio
  Astronomy, Postbus 2, 7990 AA, Dwingeloo, The Netherlands}
\altaffiltext{11}{Astronomical Institute ``Anton Pannekoek,''
  University of Amsterdam,
Science Park 904, 1098 XH Amsterdam, The Netherlands }
\altaffiltext{12}{Astro Space Center of the Lebedev Physical
  Institute, Profsoyuznaya str. 84/32, Moscow 117997, Russia}
\altaffiltext{13}{Also adjunct at the National Radio Astronomy Observatory, Green Bank, WV 24944}
\altaffiltext{14}{Department of Physics and Astronomy, University of
  British Columbia, 6224 Agricultural Road, Vancouver, BC V6T 1Z1,
  Canada} 

%\slugcomment{DRAFT: \today}

\begin{abstract}
The energetic, eclipsing millisecond pulsar J1816+4510 was recently
discovered in a low-frequency radio survey with the Green Bank
Telescope. With an orbital period of 8.7\,hr and minimum companion
mass of 0.16\,$M_\odot$ it appears to belong to an increasingly
important class of pulsars that are ablating their low-mass
companions.  We report the discovery of the $\gamma$-ray counterpart
to this pulsar, and present a likely optical/ultraviolet counterpart
as well.  Using the radio ephemeris we detect pulsations in the
unclassified $\gamma$-ray source \fgl, implying an efficiency of $\sim
25$\% in converting the pulsar's spin-down luminosity into
$\gamma$-rays and adding \psr\ to the large number of millisecond
pulsars detected by \fermi. The likely optical/UV counterpart was
identified through position coincidence ($<0\farcs1$) and unusual
colors.  Assuming that it is the companion, with $R=18.27\pm0.03$\,mag
and effective temperature $\gtrsim 15$,000\,K it would be among the
brightest and hottest of low-mass pulsar companions, and appears
qualitatively different from other eclipsing pulsar systems.  In
particular, current data suggest that it is a factor of two larger
than most white dwarfs of its mass, but a factor of four smaller than
its Roche lobe.  We discuss possible reasons for its high temperature
and odd size, and suggest that it recently underwent a violent episode
of mass loss.  Regardless of origin, its brightness and the relative
unimportance of irradiation make it an ideal target for a mass, and
hence a neutron star mass, determination.
\end{abstract}


\section{Introduction}
\begin{figure}
% plot_swift.m
%\plotone{psrj1816_allorbital.eps}
%\plotone{psrj1816_allorbital_crts.eps}
\plotone{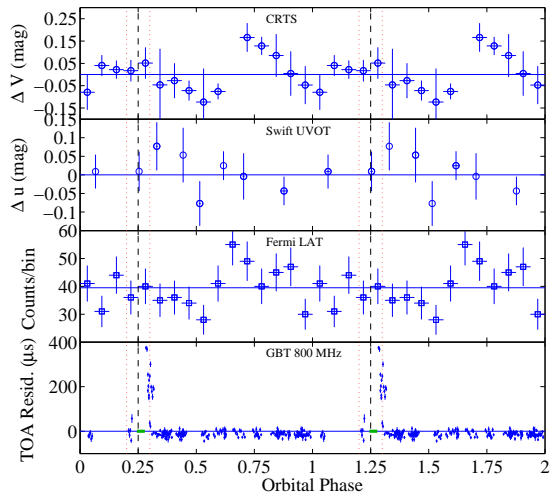}
\caption{Orbital behavior of the radio times-of-arrival (residuals at
  820\,MHz in $\mu$s from Stovall et al.\ 2012; bottom), \fermi\ LAT 0.3--10\,GeV
  count-rate (second from bottom),  \swift\ UVOT $u$-band
  photometry (relative to the mean; second from top), and CRTS
  $V$-band photometry (relative to the mean; top) versus orbital
  phase, repeated twice for clarity.  The horizontal lines are for
  reference at the mean magnitude, count-rate, and at 0 residual.  The
  \swift\ data have a scatter consistent with the uncertainties, with
  $\chi^2=5.4$ for 7 degrees-of-freedom relative to a constant
  model. For the \fermi\ data, $\chi^2=20.9$ for 15
  degrees-of-freedom. The CRTS data may show some orbital modulation
  ($\chi^2=29.2$ for 14 degrees-of-freedom) that shares some of the
  shape of the \fermi\ data, but that needs to be confirmed.  The
  vertical dashed lines are the times of conjunction (eclipse), while
  the vertical dotted lines show the approximate observed limits on
  the eclipse duration.  In the radio panel, the thick green segments
  show the phase region where we have observations but the source was
  not detected.}
\label{fig:timephot}
\end{figure}

The 3.2-ms pulsar \object[PSR J1816+4510]{J1816+4510} was discovered
as part of the Green Bank North Celestial Cap (GBNCC) survey (Stovall
et al.\ 2012, in prep.)  --- a survey of the sky north of declination
$+38\degr$ at 350\,MHz with the 100-m Robert C.~Byrd Green Bank
Telescope --- in a pointing selected for being coincident with an
unclassified \fermi\ $\gamma$-ray source (a successful search
strategy, as shown in \citealt{hrm+11}, \citealt{rrc+11},
\citealt{cgj+11}, \citealt{kjr+11}, \citealt{kcj+12}, and
others). Shortly after discovery, it was realized that the radio data
showed evidence for acceleration in an $8.66\,$hr circular orbit with
eclipses lasting up to 10\% of the orbit at 350\,MHz
(Fig.~\ref{fig:timephot}).  Eclipsing millisecond pulsars, especially
those with $\gamma$-ray counterparts, are often associated with
``black-widow'' or ``redback'' systems.  These systems harbor low-mass
companions ($\lesssim 0.05M_\odot$ for black-widows and $\sim 0.2
M_\odot$ for redbacks; \citealt*{fst88,dpm+01,asr+09}) and have been
discovered with increasing frequency in recent years (see
\citealt{roberts11} for a recent review), often in globular clusters.
The eclipses are typically long (they can cover most of the orbit;
\citealt{asr+09,hrm+11}), implying eclipsing regions larger than the
Roche lobes of the companions, and there are regions of the orbit
where the pulsar is seen through ionized plasma that delays the pulses
compared to the expected ephemeris.  The basic model for these sources
is one in which the energetic wind from the pulsar irradiates and
ablates the companion, leading to long eclipses from ionized material
in the systems \citep[e.g.,][]{sbl+01}. The companions are usually
tidally distorted, filling a significant fraction of their Roche lobes
\citep{rcf+07}, which along with heating from the pulsar's wind leads
to significant (often $>3\,$mag at wavelengths of around 5000\,\AA) optical modulation. Such systems are
interesting both because they provide a probe of the interaction
between the pulsar's wind and the companion and, ultimately, because
such systems allow measurement of neutron star masses through binary
modeling \citep*{vkbk11}.

Here we report on new and archival optical and ultraviolet data on the
counterpart of \psr.  We use these data, along with archival
$\gamma$-ray data from the \fermi\ spacecraft, to constrain the nature
of the \psr\ system.  The outline of this paper is as follows: we
first describe the archival optical and ultraviolet data that we used
to identify the counterpart to \psr\ (\S~\ref{sec:arch}), and then we
discuss new optical data from the Wisconsin Indiana Yale NOAO
telescope (\S~\ref{sec:new}).  We fit the optical/ultraviolet spectral
energy distribution in \S~\ref{sec:sed}.  We then discuss the
\fermi\ $\gamma$-ray data (\S~\ref{sec:fermi}) and the \swift\ X-ray
upper limits (\S~\ref{sec:swift}).  Finally, we discuss the
implications of our data in \S~\ref{sec:disc}, and conclude in
\S~\ref{sec:conc}.

\subsection{System Parameters}
We make use of the radio ephemeris for \psr\ determined by Stovall et
al.\ (2012): position $\alpha=18^{\rm h}16^{\rm m}35\fs9314(2)$
$\delta=+45\degr10\arcmin33\farcs864(2)$ (J2000), binary period
$P_b=8.66\,$hr, and minimum companion mass $M_c=0.162 M_\odot$
(assuming a neutron star mass of $1.4M_\odot$, although a somewhat
more massive neutron star may be likely given likely accretion
histories; \citealt{vbvs+08,vkbk11}).  Such a companion mass would put
it among the ``redback'' class \citep{roberts11}, although if the
companion is more degenerate it might instead contain a He-core white
dwarf such as that in the \object[PSR J1911-5958A]{PSR~J1911$-$5958A}
system \citep{bvkkv06}.  If the orbit is edge-on, then the full semi-major
axis is $a=2.46 R_\odot$ with a Roche-lobe radius $R_L=0.53 R_\odot$
(based on \citealt{eggleton83}).  The dispersion-measure (DM) distance
is 2.4\,kpc \citep[][for a DM of $38.8\,{\rm cm^{-3}\,pc}$]{cl02},
although given the high Galactic latitude ($b=24.7\degr$) the
uncertainties are large and this could be an underestimate
\citep{gmcm08,cbv+09,roberts11}.  Therefore we approximate the
distance as 2\,kpc and parameterize it as $d=2d_2\,$kpc, with a
nominal value of $d_2=1.2$. In what follows, other fundamental
parameters for \psr\ that are not explicitly cited are based on
Stovall et al.\ (2012).

\section{Optical and Ultraviolet Data and Analysis}
\subsection{Archival Optical/UV Data}
\label{sec:arch}
We initially identified a potential counterpart to \psr\ in the
USNO-B1.0 survey \citep{mlc+03}: the star 1351-0294859 is at $18^{\rm
  h}16^{\rm m}35\fs93$, $+45\degr10\arcmin34\fs2$.  This is $0\farcs4$
from the radio position, reasonably consistent with typical
astrometric accuracy from the USNO catalog.  However, the mean
epoch of those data is 1974, so a small proper motion could also
account for some of the difference.  The photometry for this source is
presented in Table~\ref{tab:phot}, where we have assumed uncertainties
of 0.2\,mag for the Digitized Sky Survey (DSS) photometry.

\begin{figure}
%\plotone{psrj1816_optical.ps}
\plotone{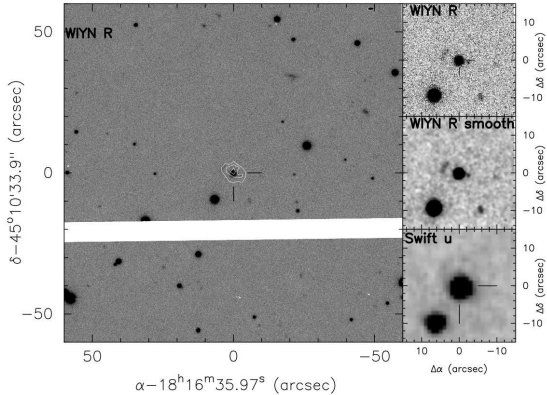}
\caption{Optical images of the \psr\ field.  The radio position is
  indicated with the tick-marks, and the uncertainties are dominated
  by uncertainties in the absolute astrometry of the optical data.
  The large image is the WIYN MiniMo $R$-band data, showing a
  $2\arcmin\times 2\arcmin$ portion.  The white band is the gap
  between the MiniMo CCDs.  The contours are from the \galex\ near-UV
  image, and the source \gal\ is consistent with a point source.  The
  insets on the right are (top to bottom): the WIYN MiniMo $R$-band
  data zoomed to show a $30\arcsec\times 30\arcsec$ portion and with a
  gray-scale adjusted to show the faint structure; the WIYN MiniMo
  $R$-band data smoothed with a Gaussian kernel with
  $\sigma=2\,$pixels ($0\farcs28$); and the \swift\ UVOT $u$-band
  image.}
\label{fig:image}
\end{figure}

The same source is identified in the Catalina Surveys Data Release 1
(CSDR1; \citealt{ddm+09}) Catalina Real-Time Transient Survey (CRTS).
The automated software actually identified \textit{two} sources that
make up the counterpart: CSS~J181635.9+451033 and
CSS~J181635.9+451036. These sources are quite close ($<2\farcs5$
apart, which is comparable to the plate-scale of the instrument) and
they both have the same average magnitude of $V_{\rm CSS}\approx
18.4$\,mag, measured with an unfiltered detector.  There were no
images where both sources were seen at the same time, and comparing
the positions of individual detections (rather than the average
positions in the catalog) it seems they are the same source that was
split by the photometric pipeline into two.  The photometry of the
combined source (106 measurements over 6.5\,yr, from 2005~May to
2011~November) is largely consistent with being constant, except for 8
points.  Three of those are clearly when the software mis-identified a
slightly brighter ($V_{\rm CSS}\approx 16.5$\,mag) star about
$10\arcsec$ to the South East (visible in Figure~\ref{fig:image}) as
being part of this object.  The others are not as easy to reject, but
since they also have $V_{\rm CSS}\approx 16.5$\,mag we think it likely
that it was another mis-identification or a photometric artifact;
without the images we cannot be certain.  Excluding those 8 points we
have data consistent with a constant $V_{\rm CSS}=18.47$\,mag with
root-mean-square variations of $0.18\,$mag.  The $\chi^2$ relative to
a constant model is slightly high (156.1 for 97 degrees of freedom),
but is similar to that for a star of similar brightness $30\arcsec$
away.  There is no evidence for any secular trends in the photometry.
The uncertainty on the mean magnitude is about 2\,mmag, but for
absolute photometry we transform from the unfiltered instrumental
magnitude to Cousins $V$ by $V=V_{\rm CSS}+0.31(B-V)^2+0.04$ with a
scatter of 0.06\,mag\footnote{See
  \url{http://nesssi.cacr.caltech.edu/DataRelease/FAQ.html}.}.  Since
this object has very nearly $B-V\approx 0$ (this assumes that the
colors are constant over time and orbital phase), we find $V=18.51\pm0.06$,
although this uncertainty may be somewhat underestimated.

We then identified the same source in the \galex\ All-sky Imaging
Survey (AIS; \citealt{2007ApJS..173..682M}) database.  Here, the
source is \object[GALEX J181635.9+451034]{\gal}\ and its position is
offset by $0\farcs4$ relative to the radio position, which is
consistent with the median offset of $1\arcsec$ found for
\galex\footnote{See
  http://galexgi.gsfc.nasa.gov/docs/galex/Documents/ERO\_data\_description\_2.htm.}.
Again, photometry is presented in Table~\ref{tab:phot}.

In the USNO source catalog, the average density of sources brighter
than $R=18.4$\,mag is $\expnt{6.1}{-4}\,{\rm arcsec}^{-2}$, so the false
association rate given the measured source offset is $\expnt{3}{-4}$,
making it very likely that we have the correct counterpart.  Moreover,
the presence of a \galex\ source at the same position with a
significant far-UV (FUV; 152\,nm) detection makes it essentially
certain: there are only 4 FUV detections with magnitudes brighter than
19.9\,mag within a $10\arcmin$ radius, so the false association rate is
$\expnt{2}{-6}$.  As we will see below, this source is so bright and
blue that it is rather unusual, making the chance of finding one
within $0\farcs5$ of the radio position by chance extremely low.

We then identified an observation with the \swift\ satellite
(Observation ID 00041440003).  We see a bright source at the radio
position in the data from the Ultraviolet and Optical Telescope (UVOT;
\citealt{rkm+05}).  The observation was on 2010~August~04, and
consisted of 8 separate integrations in the $u$ filter (central
wavelength of 3450\,\AA) spread over 11\,hours with a
total integration of 3173\,s and $2\times 2$\,pixel binning.  We
determined both time-resolved and summed photometry from these data
using \swift\ data-reduction tools.  First, we ran the task
\texttt{uvotsource} on each separate observation (along with
respective exposure maps) to measure how bright the object was in each
individual integration (with the 2011~October~31 calibration
database).  We then summed the integrations using \texttt{uvotimsum}
and measured the summed magnitude using \texttt{uvotsource}, where in
both cases the source region was a circle with $5\arcsec$ radius
centered on the radio position and the background region is
$25\arcsec$ in radius centered near the pulsar but not including any
visible sources.  The best-fit position of the source is $18^{\rm
  h}16\arcmin35\farcs93$, $+45\degr10\arcmin34\farcs0$, or $0\farcs12$
away from the radio position.  This is without any additional
boresight correction beyond that applied by the \swift\ processing.
The final detection significance in the summed image was
72.8\,$\sigma$, but our photometry includes the suggested systematic
uncertainty of 0.02\,mag in addition to the statistical uncertainty.

In Figure~\ref{fig:timephot} we show the measured \swift\ and CRTS
photometry as a function of orbital phase, where the
\swift\ observation times have been corrected to the Solar System
barycenter using the \texttt{barycorr} task, and the CRTS observation
times have been corrected to the heliocenter using the \texttt{rvcorr}
task in \texttt{IRAF} (for an 8-hr orbit, the differences between
helio- and bary-center are negligible).  The \swift\ data are
consistent with no variations, with $\chi^2=5.4$ for 7
degrees-of-freedom.  Each individual measurement can largely be
considered instantaneous, as the exposure times are at most 800\,s
compared to an observation duration of 11\,h and an orbital period of
8.7\,h. The rms scatter of the data about the mean is 0.05\,mag.  The
CRTS data have been binned, with between 2 and 16 individual
observations averaged into each point.  There may be a trend with
orbital phase, such that the data are slightly (15\%) fainter near
phases\footnote{Note that for pulsars ephemerides, conjunctions occur
  at phases of 0.25 and 0.75, with 0.75 having the pulsar between the
  companion and the observer.} of 0.75, but while formally significant
($\chi^2=29.2$ for 14 degrees-of-freedom) we do not know the level of
systematic uncertainties due to artifacts and mis-identifications.
There were no measurements in the bin just before the apparent flux
minimum, but without the raw data we cannot say whether there were no
observations or just no detections.  The rms scatter of the data about
the mean is 0.08\,mag.  The possible trend in the CRTS data is not
necessarily seen in the \swift\ data, although it is difficult to be
certain.

\subsection{New Optical Data}
\label{sec:new}
We observed \psr\ using the Mini-Mosaic Imager (MiniMo) on
the 3.5-m Wisconsin Indiana Yale NOAO (WIYN) telescope \citep{sasc00}.
The data were a 300\,s exposure in the Harris-$R$ filter on
2011~March~24 taken shortly before sunrise.  Seeing was $0\farcs7$,
with a plate-scale of $0\farcs14\,{\rm pixel}^{-1}$.  The data were
corrected for bias level and flatfielded using standard procedures in
\texttt{MIDAS}.  The image was registered to the International
Coordinate Reference Frame (ICRF) using 130 Two-Micron All-Sky Survey
\citep[2MASS;][]{2mass} stars, giving fits with rms residuals of
$0\farcs2$ in each coordinate.  We did photometric calibration using
an observation of the Landolt 98 field \citep{stetson00} earlier in
the night, determining a zero-point using 22 stars; we estimate a
zero-point uncertainty of 0.03\,mag.  As in the other data, there was
a bright source at the position of the pulsar, but here the position
offset was only $0\farcs02$.  We measured the object using
\texttt{sextractor} \citep{ba96}, with the same settings that we used
for the standard stars.  In addition to the potential counterpart, we
see some faint structure to the north-east and south-west, at
distances of $2\arcsec$ to $4\arcsec$.  This may just be some faint
stars near the detection limit of the image, but they are also
somewhat suggestive of an H$\alpha$ bowshock nebula such as that seen
around the black-widow system \object[PSR B1957+20]{PSR~B1957+20}
\citep{kh88} or around the non-interacting pulsar/white dwarf binary
\object[PSR J0437-4715]{PSR~J0437$-$4715} \citep{bbm+95}.  Images of
both the WIYN and \swift\ data are shown in Figure~\ref{fig:image}.

\subsection{Optical/Ultraviolet Spectral Energy Distribution}
\label{sec:sed}
\begin{figure}
%\plotone{psrj1816_UmR_R.eps}
\plotone{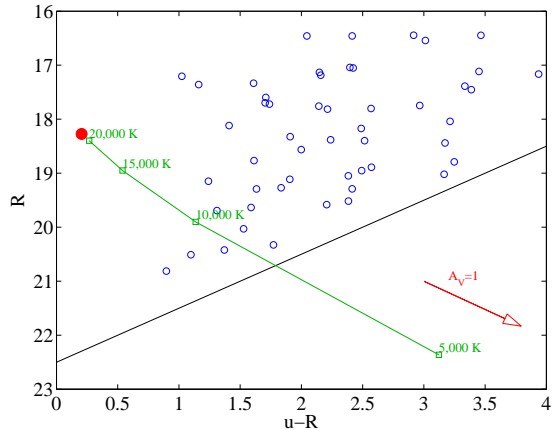}
\caption{Color-magnitude diagram for the sources in the WIYN $R$-band
  and \swift\ $u$-band images.  The source at the position of the
  pulsar is the filled circle.  The black line is the approximate
  $u=22.5$ detection limit of the UVOT image.  We show the results of
  a reddened ($A_V=1\,$mag) blackbody model for the temperatures
  shown, with a size of $0.005R_\odot/100\,$pc.  A reddening vector showing
  1\,magnitude of extinction is also shown.}
\label{fig:cmd}
\end{figure}

The potential counterpart of \psr\ is very blue compared to nearby
sources.  In Figure~\ref{fig:cmd} we show a color-magnitude diagram
using the $R$-band and $u$-band data, where the counterpart is roughly
1 magnitude bluer than the field sources.  Just the $u$ and $R$ data
indicate a rather hot blackbody, although since the reddening vector
is roughly parallel to the track of a blackbody the temperature and
reddening are degenerate.  However, some temperature above 10,000\,K
is required.

We fit all of the optical/UV photometry to determine the spectral
energy distribution (SED) of this source.  For the data in the Vega
system, we used zero-point fluxes from \citet*{bcp98}.  We then
convolved various model SEDs with filter transmission curves and
compared the resulting fluxes with those derived from the data.  For
the DSS and CSS data we assumed standard Johnson $BRI$ filters, and
this is clearly a simplification, but the large uncertainties for DSS
mean that those data have modest weight.  For the WIYN data we used a
filter curve from \citet{kfwa09}; for the \swift\ data we used a
response file from the \swift\ web
site\footnote{\url{http://heasarc.nasa.gov/docs/swift/proposals/swift\_responses.html}.};
the \galex\ filters were from the COSMOS web
site\footnote{\url{http://www.astro.caltech.edu/{\til}capak/cosmos/filters/}.}.
We used a nominal extinction curve from \citet*{ccm89} and \citet{o94},
with a reddening ratio $R_V=3.1$.

\begin{figure}
%\plotone{psrj1816_optical_fit.eps}
\plotone{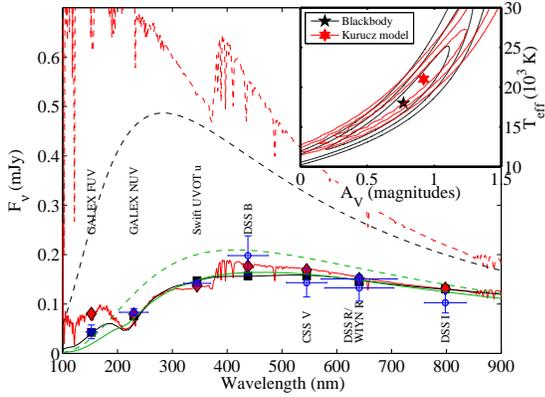}
\caption{The results of fitting to the broadband photometry from
  Table~\ref{tab:phot}.  We plot the best-fit SEDs, with the blackbody
  model (black) and model atmosphere (red).  The solid curves have had
  extinction applied, while the dashed curves are without extinction.
  The black squares and red diamonds are the respective model SEDs convolved with
  the filter passbands, while the circles are the data.  The inset
  shows contours of $\chi^2$ vs.\ extinction $A_V$ and effective
  temperature $T_{\rm eff}$ for the blackbody models (black) and model
  atmospheres (red).  Contours are at 1-, 2-, and 3-$\sigma$
  confidence on the joint fit.  The stars represent the best-fit
  parameters for each model. For comparison we also plot the best-fit
  blackbody with $A_V=0.2\,$mag (green).}
\label{fig:sed}
\end{figure}

Our first model was a reddened blackbody.  We got a good fit, with
$\chi^2=6.4$ for 5 degrees of freedom (and this includes the poorly
calibrated DSS data).  The best-fit model had $T_{\rm eff}=18$,000\,K,
$A_V=0.77\,$mag, and size $R=0.1R_\odot$ at a nominal
distance of 2\,kpc, but a wide range of solutions had similarly good
fits (Figure~\ref{fig:sed}) with larger temperature requiring larger
extinctions and smaller radii; as in Figure~\ref{fig:cmd} this is
largely the result of the blackbody model between the $u$ and $R$ data
being parallel to the reddening vector.  The size is largely
determined from the WIYN observation, with
%\begin{eqnarray}
\[
\log_{10} \left[\left(\frac{R}{R_\odot}\right)\left(\frac{2\,{\rm
      kpc}}{d}\right)\right]  \approx  -0.54-0.35
\log_{10}\left(\frac{T_{\rm eff}}{10^3\,{\rm K}}\right)
\]
\[
A_V \approx  -3.09 \log_{10}\left(\frac{T_{\rm eff}}{10^3\,{\rm
    K}}\right)^2+10.65\log_{10}\left(\frac{T_{\rm eff}}{10^3\,{\rm
    K}}\right)-7.72 
\]
along the best-fit locus.

We then fit model stellar atmospheres from \citet{kurucz93}.  We used
models with gravity $10^4\,{\rm cm\,s}^{-2}$, although the results
were not sensitive to this.  The best-fit model was slightly hotter
than the best-fit blackbody (21,000\,K, with $A_V=0.92\,$mag and
$R=0.12R_\odot$ at a distance of 2\,kpc), and had a slightly worse
$\chi^2$ (11.5), but uncertainties in the extinction law below 250\,nm
\citep{ccm89} could change the result; the most discrepant point was
the \galex\ FUV observation.  The best-fit region in the ($A_V,T_{\rm
  eff}$) plane is very similar to that of the blackbody
(Figure~\ref{fig:sed}).  We also tried a 15,000\,K white dwarf
atmosphere model (a DA white dwarf with hydrogen on the surface, which
is expected for such a hot star; \citealt{hl03}), kindly supplied by
D.~Koester.  The fit was similar to that of a 15,000\,K main-sequence
star since the main differences (the stronger Balmer absorption
sequence in the white dwarf) are not easily distinguished with the
available DSS $U$-band photometry; the white dwarf model also had
trouble fitting the \galex\ FUV point.

An absorbed power-law spectrum such as that of an active galactic
nucleus does not fit ($\chi^2=34.7$ for 5 degrees of freedom).  While
the spin-down and radio emission of \psr\ indicate that no accretion
disk is likely present \citep[cf.][]{asr+09}, given the unusual nature
of the optical/UV emission it may be worth considering whether the
high temperature that we measure is from an accretion disk around the
pulsar, where ultraviolet emission is common (although this would tend
to also have bright X-ray emission which we do not see, the X-rays
could be variable).  %The Roche radius for the pulsar is $1.3R_\odot$,
%and the emission we see likely comes from a much smaller area unless
%it is at a high inclination where the cool edges of the disk obscure
%the hot inner portion. 
The SED that we measure is consistent with a
single-temperature blackbody, not a multi-temperature model as is
usually used to model accretion disks \citep{ss73,vrg+90}.  Therefore,
while we cannot rule out such a model using photometry alone, we
consider it unlikely.  Spectroscopy should be definitive.

We note that the faint, diffuse emission seen in the $R$-band image
could have contaminated some of the lower-resolution data (in
particular \swift\ and \galex).  However, at least at $R$-band the
brighter spot (to the south-west) is about 100 times fainter than the
star, so assuming a similar spectrum (which is conservative for the
bluer bands, as the emission is probably either Balmer dominated or
stellar) it is unlikely to be important at the $>1$\% level.  We tried
subtracting the stellar point-spread-function (PSF) at the position of
the counterpart, but do not see any significant residual emission
beyond that visible in Figure~\ref{fig:image}.

\section{X-ray and $\gamma$-ray Data and Analysis}
\label{sec:highenergy}
\subsection{\textit{Swift} X-ray Data}
\label{sec:swift}
In the 2.8\,ks photon-counting observation with X-ray Telescope (XRT;
\citealt{bhn+05}) 0 counts were detected in a circle with radius
$20\arcsec$ around the radio position.  Millisecond pulsars typically
have a combination of X-rays from thermal (from hot polar caps) and
non-thermal (either magnetospheric, or from a shocked pulsar wind)
spectra \citep[e.g.][]{zavlin07}.  The thermal components have
blackbody temperatures with $kT\approx 0.1$--0.2\,keV and luminosities
of $\sim 10^{-3} \dot E$ \citep{zavlin07,pccm02}.  For more energetic
pulsars the non-thermal components dominate, and these are typically
fit as power-laws with photon indices $\Gamma\approx 1.5$ and again
luminosities of $\sim 10^{-3} \dot E$.  Based on a power-law with
photon index of 1.5, we set a 2\,$\sigma$ limit of
$\lesssim\expnt{3}{-15}\,{\rm erg\,s}^{-1}\,{\rm cm}^{-2}$ for the
unabsorbed 0.5--10\,keV flux from \psr\ for absorption column
densities $N_{\rm H}$ in the range of $10^{20}\,{\rm
  cm}^{-2}$--$10^{21}\,{\rm cm}^{-2}$ (corresponding to
$A_V=0.06-0.6\,$mag, based on \citealt{ps95}).  For a measured
spin-down luminosity of $\dot E=\expnt{5}{34}\,{\rm erg\,s}^{-1}$
(although see below for possible corrections to this), our limit then
corresponds to $L_{\rm X,non-th}\lesssim\expnt{3}{-4} d_2^{-2} \dot
E$.  We can do the same computation for a thermal spectrum, with
unabsorbed flux limits of $\expnt{(9-17)}{-15}\,{\rm
  erg\,s}^{-1}\,{\rm cm}^{-2}$ (0.5--2\,keV) for a blackbody with
$kT=0.15\,$keV and $N_{\rm H}=\expnt{(1-10)}{20}\,{\rm cm}^{-2}$. This
then gives a similar limit of $L_{\rm X,thermal}\lesssim\expnt{2}{-4}
d_2^{-2} \dot E$.  Both of these efficiencies are low but not outside
the observed range \citep{zavlin07,rrc+11}, and suggest that the X-ray
flux may only be slightly below the \swift\ limit.

\subsection{\fermi\ Data}
\label{sec:fermi}
The radio position of \psr\ matches almost exactly with a source from
the \fermi\ Large Area Telescope Second Source Catalog (2FGL;
\citealt{2fgl}).  The source \object[2FGL J1816.5+4511]{\fgl}
(1FGL~J1816.7+4509 from the first year catalog) is $1.4\arcmin$ away
from \psr, with a position uncertainty of $\approx 5\arcmin$ in
radius.  It is listed as having a power-law spectrum with photon index
$\Gamma=2.11\pm0.08$ ($N_E\propto E^{-\Gamma}$), and 0.1--100\,GeV
flux of $\expnt{(15.3\pm1.8)}{-12}\,\ergscm$.

However, there is another year of data available in the
\fermi\ archive, and we wished to do further spectral analysis and
look for pulsations.
Therefore we analyzed data from the \fermi\ Large Area Telescope (LAT;
\citealt{aaa+09b}), including events from 2008~August~05 to
2012~January~19.  We followed standard procedures\footnote{See
  \url{http://fermi.gsfc.nasa.gov/ssc/data/analysis/scitools/data\_preparation.html}.}
in filtering events, selecting those with event class 2 within a
$10\degr$ radius around \psr, with energies between 0.2--10\,GeV (to
avoid the poor point-spread function at the lowest energies; the
pulsar did not appear to be detected there anyway) and zenith angles
$<105\degr$.  We computed the spectrum using an unbinned likelihood
analysis\footnote{See
  \url{http://fermi.gsfc.nasa.gov/ssc/data/analysis/scitools/likelihood\_tutorial.html}.},
including the contributions of sources out to a radius of $17\degr$
from the 2FGL catalog as well as isotropic and Galactic background
models appropriate for pass 7 data (models
\texttt{iso\_p7v6source.txt} and \texttt{gal\_2yearp7v6\_v0.fits})
with the \texttt{P7SOURCE\_V6} instrument response, although we held
most of the source parameters fixed at their catalog values with the
exceptions of those sources within $8\degr$ of \psr\ and the diffuse
background normalizations.  Photons are significantly detected between
0.5\,GeV and 5\,GeV.  For \fgl\ we find a good fit with a power-law
model with photon index $\Gamma=2.20\pm0.07$ and normalization
$\expnt{(1.5\pm0.1)}{-12}\,\cmsMeV$ at 1.15\,GeV giving an integrated
0.1--100\,GeV flux of $\expnt{(19.6\pm1.5)}{-12}\,\ergscm$, with a
Test Statistic of 404 (i.e., roughly a $20\,\sigma$ detection); this
is similar to the result from the 2FGL catalog.  We do not incorporate
any systematic uncertainties related to calibration errors.

We repeated the spectral fit with a power-law modified by an
exponential cutoff, $N_E\propto E^{-\Gamma}e^{-E/E_c}$.  The meager
energy range with significant detections meant that the cutoff could
not be strongly constrained, but we find $\Gamma=2.0\pm0.1$,
normalization $\expnt{(7.9\pm0.8)}{-12}\,\cmsMeV$ at 0.55\,GeV, and
cutoff energy $E_c=7.5\pm4.0$\,GeV; the integrated flux was
$\expnt{(15\pm3)}{-12}\,\ergscm$.  Formally this fit was
statistically equivalent to the pure power-law fit, and other local
minima were also possible depending on where the fit was started.  We
note that the parameters we find are outside the range of most
millisecond pulsars, with $\Gamma$ and $E_c$ both higher than are
typical.  Much of this comes from the highest energies we included in
our fit (4--6\,GeV; Fig.~\ref{fig:fullsed}).  Without this bin, a fit
with more typical values ($\Gamma\approx 1.5$, $E_c\approx 3\,$GeV) is
acceptable.  We show the fits in Figure~\ref{fig:fullsed}, where
$\gamma$-ray fluxes were determined from modeling the flux in each
energy bin as a single power-law\footnote{See
  \url{http://fermi.gsfc.nasa.gov/ssc/data/analysis/scitools/python\_tutorial.html}.}
using the contributed task \texttt{likeSED}.  We therefore urge
caution in interpreting the $\gamma$-ray spectrum.

\begin{figure}
%\plotone{psrj1816_fermipulse.eps}
% plot_fermipulse_scottmew.m
%\plotone{psrj1816_fermipulse_radio}
\plotone{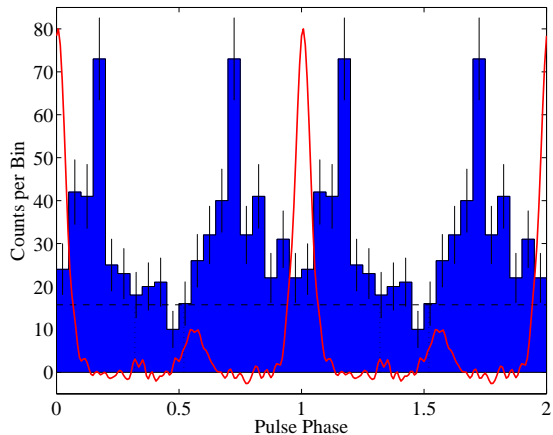}
\caption{The pulsed $\gamma$-ray light curve of \fgl, determined from
  all LAT data in the 0.3--10\,GeV range up to 2012~January and
  repeated twice for clarity.  We used the radio ephemeris to
  initially identify the pulsar, but then refined the ephemeris using
  the \fermi\ data since they have a longer time span.  The dashed
  horizontal line is the approximate background level based on the
  phases indicated by the vertical dotted lines.  The red trace is the
  radio pulse profile, based on 820-MHz data from the Green Bank
  Telescope (Stovall et al.\ 2012), arbitrarily scaled.}
\label{fig:pulse}
\end{figure}

For the pulsation search, after the initial event filtering, we
assigned phases to all of the events using the best-fit radio
ephemeris using the \fermi\ plugin for \texttt{tempo2}\footnote{See
  \url{http://fermi.gsfc.nasa.gov/ssc/data/analysis/scitools/pulsar\_analysis\_appendix\_C.html}.}
\citep*{hem06}.  We detected pulsations using the initial radio
ephemeris, but given the longer time span of the \fermi\ data compared
to the radio data (1262\,days vs.\ 320\,days) we were able to refine
the radio ephemeris (in particular the spin-down), as discussed in
Stovall et al.  We then used the refined ephemeris to update the event
phases.  Selecting the {632} events $\leq 0.65\degr$ from the radio
position and with energies between 0.3\,GeV and 10\,Gev
\citep[optimizing those parameters for the pulse amplitude, as
  in][]{rrc+11}, we see very significant pulsations, with a 
$H$-test statistic \citep*{drs89} of {64.4 for 11 harmonics (false
  alarm probability of $\expnt{4}{-23}$)}.
With this solution we see clear pulsations in the binned light curve
with $\chi^2=157$ for 19 degrees-of-freedom.  The pulsations have two
sharp peaks separated by slightly less than one half of a cycle
(Fig.~\ref{fig:pulse}) similar to the radio pulse.  Selecting events
from phases 0.32--0.52 (where the pulse is at a minimum), we see a
radial profile that is consistent with being flat in terms of counts
per unit area out to $2\degr$, with an implied background rate over
all phases of $237\pm38\,{\rm deg}^{-2}$ (the horizontal line in
Fig.~\ref{fig:pulse}). This would mean that our lightcurve is
consistent with being 100\% pulsed.  Averaging over pulse phase, the
count-rate was roughly constant as a function of orbital phase
(Fig.~\ref{fig:timephot}), with $\chi^2=20.9$ for 15
degrees-of-freedom.  Over the two years of the 2FGL catalog the
lightcurve was likewise consistent with being constant on month
timescales ($\chi^2=22.1$ for 23 degrees-of-freedom, based on the 2FGL
variability index).

\section{Discussion}
\label{sec:disc}
Based on the \fermi\ flux of $\expnt{20}{-12}\,{\rm erg\,cm}^{-2}\,{\rm
  s}^{-1}$ (0.1--100\,GeV) we find a $\gamma$-ray luminosity of
$\expnt{1}{34}d_2^2\,{\rm erg\,s}^{-1}$.  We determine the
$\gamma$-ray efficiency by comparing this with $\dot E$ and find
$\eta_\gamma\equiv L_\gamma/\dot E=0.19 d_2^2$ assuming no geometric
beaming correction; this is comparable with that found for purely
magnetospheric emission from millisecond pulsars with
\fermi\ \citep{aaa+09,rrc+11}.  We can set a weak upper limit to the
distance by requiring $L_\gamma\leq \dot E$, which gives $d_2\lesssim
2.3$ ($d\lesssim 4.6\,$kpc).  Given the highly pulsed $\gamma$-ray
lightcurve, it is possible that all of the emission could be
magnetospheric in origin like for most millisecond pulsars, but some
might still be related to intra-binary shocks such as in the
\object[PSR B1259-63]{PSR~B1259$-$63} system \citep[][which is not a
  millisecond pulsar but has a hot companion like that seen
  here]{aaa+11}, especially if there are contributions from inverse
Compton scattering off the UV photons present in both cases.

\begin{figure}
%\plotone{psrj1816_fullsed.eps}
\plotone{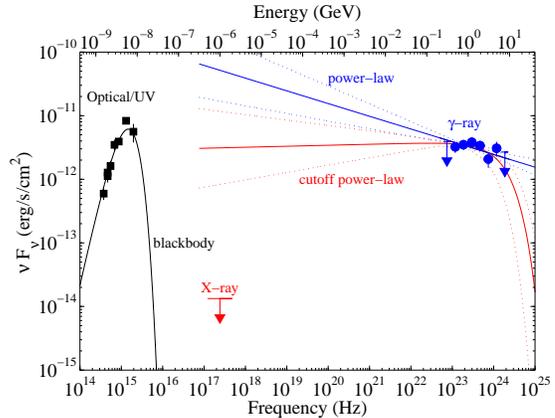}
\caption{The SED of \psr\ and its presumed companion, from the optical
  to $\gamma$-rays.  We show the optical/UV photometry along with the
  best-fit blackbody model, both corrected for reddening with
  $A_V=0.2\,$mag.  The X-ray upper limit is based on the \swift\ XRT
  non-detection.  The $\gamma$-ray points were derived from fitting a
  power-law model to each individual energy bin, and we show the
  best-fit single power-law and cutoff power-law models along with
  $\pm1\,\sigma$ uncertainties.}
\label{fig:fullsed}
\end{figure}

In Figure~\ref{fig:fullsed} we plot the spectral energy distribution
from optical to $\gamma$-rays. Energetically, the optical/UV are
almost as important as the $\gamma$-rays, which would make it
difficult for them to both ultimately come from $\dot E$ (as the
$\gamma$-rays already require a substantial fraction of $\dot E$),
supporting a hot companion which radiates on its own.  Changing the
extinction to a lower value such as $A_V=0.2\,$mag (see below) reduces
the total optical luminosity somewhat, but it is still substantial.
The simple power-law fit to the $\gamma$-rays exceeds the X-ray upper
limit (much as in \citealt{dkp+11}), but the cutoff power-law does as
well, so with either model we might need a spectral break between
1\,keV and 100\,MeV.  However, our spectral fitting only had a limited
number of counts, and we did not include systematic uncertainties
related to instrumental calibration.  The apparent discrepancy between
the $\gamma$-ray emission and the X-ray upper limit may only be a
consequence of the spectral fit; the ratio of $\sim 10^3$ between the
$\gamma$-ray and X-ray luminosities is reasonable for other
millisecond pulsars (\citealt{rrc+11}; \citealt*{tct12}).

In what follows, we primarily assume that the optical/UV emission come
from a single photosphere that is the companion of the pulsar in a
binary system.  However, without a spectrum we cannot exclude
contributions from shocked plasma --- this might help explain some of
the slightly discrepant UV data-points or the somewhat high extinction
(see below).

Given our extremely likely detection of the optical/UV companion of
\psr, and the identification of the pulsar at  $\gamma$-ray energies, we
consider how this system fits among the known recycled pulsars with
low-mass companions.  Some of the implications of this system are
common to a wide range of similar systems, but the uniquely hot
temperature may point to either a fortuitous detection of a
short-lived evolutionary state or a different evolutionary path.

At the nominal distance, our optical photometry implies $R=0.1 d_2
R_\odot$ ($L_{\rm optical/UV}\approx L_\odot$).  If the optical
companion filled its Roche lobe, it would be at a distance of 10\,kpc,
which is not impossible given the radio and $\gamma$-ray properties
but is unlikely (it would require $L_{\gamma}\approx 4\dot E$, but
this constraint is based on isotropic emission).  If the optical
source is an unrelated object, it would either be a nearby white dwarf
at $\lesssim 300\,$pc (\citealt*{fbb01}\footnote{Also see
  \url{http://www.astro.umontreal.ca/{\til}bergeron/CoolingModels/}.};
we take it as a $0.6M_\odot$ carbon/oxygen white dwarf) or a main
sequence star (B3--5) at a distance of 70--100\,kpc \citep{allen};
while the former is possible, main sequence stars with $M\gtrsim
3M_\odot$ would not be expected at such distances.

Our best-fit value for the extinction $A_V$ is 0.5--1.0\,mag.  This is
larger than the largest value in this direction ($\approx 0.2\,$mag)
determined by \citet*{dcllc03}, and \citet*{sfd98} give a similar
result.  Given the limitations of our fit, values of $\lesssim
0.2\,$mag are not excluded but would imply effective temperatures of
$\approx 12$,000\,K ($A_V=0.2\,$mag increases $\chi^2$ by 2.2 over the
best-fit value for the blackbody fit), and we plot the best-fit
$A_V=0.2\,$mag blackbody for comparison in Figure~\ref{fig:sed}.  To
evaluate the likely extinction, we determined our own run of
extinction with distance by examining all of the 2MASS stars within
$1\degr$ and finding the red clump \citep{dcllc03,dvk06}.  While we
cannot determine the extinction as close as 2\,kpc (there are not
enough stars), we measure extinctions of $\approx 0.6\,$mag for
distances $\geq 5\,$kpc, which is reasonably consistent with our SED
fitting.  At this Galactic latitude ($b=+24.7\degr$) much of the
extinction will be close to the Sun, so the value measured at 5\,kpc
should be applicable to closer objects.  We note that our value of
$A_V\gtrsim 0.5\,$mag is actually consistent with the measured DM,
assuming an ionized fraction of 10\% and the usual \citep{ps95}
conversion between extinction and hydrogen column density, but the
hydrogen column density interpolated from \ion{H}{1}
maps\footnote{Using \url{http://asc.harvard.edu/toolkit/colden.jsp}.}
is lower, $\expnt{4}{20}\,{\rm cm}^{-2}$.  Optical and X-ray
spectroscopy can hopefully narrow down the possible range of the
extinction.
% about 70\% of the total electron density in the \citet{cl02} model
%for this line of sight, suggesting

%\textbf{Should this be in this paper?}\\ 
With an minimum eclipse duration of 7\% of the orbit at 820\,MHz
(Stovall et al.\ 2012), the eclipsing radius is $\approx 0.5R_\odot$.
This is similar to the Roche lobe radius, suggesting that some of the
eclipsing material may be gravitationally bound to the companion star,
although the tail of delayed times-of-arrival (TOAs) extends to larger
radii.  The maximum delay observed at 820\,MHz (about 400\,$\mu$s)
implies an excess dispersion measure of $0.06\,{\rm pc\,cm}^{-3}$, or
an electron column density of $N_e\approx \expnt{2}{17}\,{\rm
  cm}^{-2}$.  If this material is distributed over $0.5 R_\odot$, we
would have an electron density $n_e\approx \expnt{6}{6}\,{\rm
  cm}^{-3}$.  Assuming that the material is moving at the escape
velocity, we estimate a mass-loss rate of $\dot M\sim
10^{-13}M_\odot\,{\rm yr}^{-1}$, so the companion would not be
substantially diminished over a Hubble time (similar to
\citealt{sbl+96}). Such mass-loss rate is actually comparable with
expectations for radiative winds from more massive sdB stars with
(presumably) similar surface gravities\footnote{Although we are
  outside their nominal luminosity range and near the low end of the
  temperature range usually considered, the surface gravities are
  similar so the winds are likely to be similar.}
\citep{vc02,unglaub08}.  If this and not ablation (which would only
require 0.1\% of $\dot E$) is the origin of the mass loss, then the
low gravity and high temperature of the companion seem to be necessary
components for the ionized gas eclipses as winds cease for gravities
$>10^6\,{\rm cm\,s}^{-2}$ and temperatures $<20,$000\,K
\citep{unglaub08}.  While minor in terms of mass loss the winds might
be nonetheless important evolutionarily in altering the diffusive
equilibrium \citep{ub98} and hence the atmospheric appearance and
onset of shell burning.

The measured effective temperature of $\gtrsim 15$,000\,K is far
hotter than the companion to any known black-widow or redback system
(typically $\lesssim 6000\,$K; \citealt{vkbjj05,pdf+10}; Breton et
al.\ 2012, in prep.; C.~Bassa 2011, pers.\ comm.) by a factor of
almost 3.  Pulsars with hot white-dwarf companions are known (e.g.,
8,550\,K for \object[PSR J1012+5307]{PSR~J1012+5307}, 10,090\,K for
PSR~J1911$-$5958A, and 15,000\,K for \object[PSR
  B0820+02]{PSR~B0820+02}; \citealt{vkbjj05} and references therein),
but they do not have broad eclipses like those we see here.  We can
then address the peculiarity of this system in two ways: (1) if this
is an interacting binary system, what would be the consequences of it being so
hot, and (2) why is it so hot.

For the first question, if we treat \psr\ as an interacting binary
regardless of origin, it is not surprising that we do not see any
modulation of the $u$-band lightcurve: the additional energy deposited
by the pulsar has an equilibrium temperature $\lesssim (\dot E/4\pi
a^2 \sigma)^{1/4}\approx 7000\,$K (for an efficiency of 100\%, while
typical efficiencies are closer to 10\%; Breton et al.\ 2012, in
prep.).  Note that this ascribes all of the observed spindown $\dot
P=\expnt{4.1}{-20}\,{\rm s\,s}^{-1}$ to magnetic dipole radiation and
not to secular (i.e., \citealt{shklovskii70}) acceleration; with a
typical millisecond pulsar velocity of $v=100 v_{100} \,{\rm
  km\,s}^{-1}$ the secular $\dot P$ would be $\expnt{1.7}{-21}
v_{100}^2 d_2^{-1} \,{\rm s\,s}^{-1}$ or 4\% of the measured value, so
it is likely that the $\dot E$ value we use above is close to correct
(corrections due to differential Galactic acceleration are even
smaller; \citealt{nt95}).
Since the implied temperature is so low compared to the observed
temperature, we would expect the illuminated side to be $\lesssim 5\%$
brighter than the dark side (for an effective temperature of
15,000\,K), which is the level of the observed scatter in the
\swift\ photometry; if the orbital modulation of the CRTS data is
real, then it is difficult to understand its amplitude.  We note that
our SED fitting assumes that the data are constant with time and are
defined by only a single model, and these are not necessarily valid
assumptions.  Even aside from the lack of strong orbital modulation,
there could also be secular/state changes such as those seen in
\object[PSR J1023+0038]{PSR~J1023+0038} \citep{asr+09}.  However, the
limit on secular evolution from the CSDR1 data (which span the times
of the \galex, \swift, and WIYN Photometry) suggests that most of the
data are consistent with a single model.  In the future, single-epoch
photometry should be able to resolve this question.  The nominal
radius of the companion is well within the typical range for
black-widow/redback companions in the field (sources in globular
clusters often are considerably $>R_\odot$;
\citealt{fpds01,cfp+08,pdf+10}), which may be some clue to the
formation mechanism.  However, unlike some of those systems, it is
likely a factor of several smaller than the Roche lobe, so the
companion may not be significantly distorted.

It may be that the $0.5 R_\odot$ eclipse duration is set not by the
Roche lobe radius but by the intrabinary shock radius between the
$\dot E$-driven wind and the companion \citep{at93}, where the shock
represents the equilibrium between the relativistic wind of the pulsar
and the presumed radiation-driven wind from the companion.  Assuming
an electron density of $\sim 10^7\,{\rm cm}^{-3}$ and pure hydrogen
composition at a radius of $R_{\rm shock}\approx 0.5 R_\odot$, the ram
pressure $\rho v^2/2$ is $0.03\,{\rm dyne\,cm}^{-2}$ for a wind at the
escape velocity, which is a factor of $\sim 200$ less than $\dot
E/(4\pi c (a-R_{\rm shock})^2)$; if this model is valid, then we must
consider that the wind might be moving faster than the escape speed
\citep{pebk88,unglaub08}, our density might be too low (in particular
the material might be clumpy), or possible only a fraction of $\dot E$
participates in the shock \citep{sgk+03}.

For the second question, we can ask why the companion would be so hot.
First, it is possible that \psr\ has a normal cool companion, but that
the optical/UV flux comes from another source.  This could be a star,
either as part of a triple system or an unrelated object.  Having an
unrelated object seems highly unlikely given the positional
coincidence, and an association between the radio, $\gamma$-ray, and
optical sources is the most likely explanation, but without optical
modulation to confirm we cannot be certain.  A triple system can
largely be ruled out by the 3-yr span of the \fermi\ timing, as those
data are consistent with only the 8-hr binary.  As mentioned before,
emission from shocked plasma is also possible.

Without knowing the surface gravity (and hence having some idea of how
degenerate the companion is), any inferences about why the companion
is so hot are difficult.  For helium-core white dwarfs in the mass
range considered here ($\lesssim 0.2M_\odot$) burning of a thick shell
of hydrogen \citep{webbink75} can keep the sources hotter for
considerably longer (Gyr; \citealt{sega00,pach07}; \citealt*{sba10})
than standard cooling would allow \citep[cf.][]{llfn95}, but the
temperatures tend to be $\lesssim 10$,000\,K; more massive white
dwarfs can stay at $>10^4\,$K for longer, and unstable hydrogen
flashes can also push the temperature above $10^4\,$K for more massive
white dwarfs temporarily.  We note, though, that the recently
discovered binary \object[SDSS
  J065133.3+284423.3]{SDSS~J065133.33+284423.3} \citep{bkh+11} has a
helium-core white dwarf at a similar temperature to what we find, and
this source is somewhat more massive ($0.25M_\odot$) than the expected
burning limit, although the limit is metallicity-dependent and the
contribution of tidal heating in this system is unknown.  If the
companion of \psr\ were a normal white dwarf we would expect \psr\ to
be on the low side of the possible distances (radii of $\lesssim
0.05R_\odot$ are typical at these masses; \citealt{pach07}) with an
age of $\lesssim 0.5\,$Gyr.  In this scenario, \psr\ would be the
first pulsar/white dwarf system with ionized-gas eclipses.  However,
it could also be that the companion is still hotter and younger still
if it has not yet equilibrated from a common-envelope evolutionary
phase \citep{paczynski76} or Roche-lobe overflow that stripped away
the outer layers of the companion, leaving it hot, large, and in a
circular orbit \citep[e.g.,][]{dsbh98}.  For low-mass
($<0.2\,M_\odot$) white dwarfs, the residual hydrogen burning
typically results in luminosities of $<0.1L_\odot$
\citep{pach07,sba10}, and therefore it might be that the luminosity we
see is dominated by gravitational contraction (but see
\citealt{dsbh98}).  If that is true, we estimate a thermal timescale
of $2d_2^{-3}\,$Myr for an effective temperature of 18,000\,K and
considering the whole star --- the timescale would be $2d_2^{-3}\,$kyr
if we only are concerned with a typical envelope of $10^{-3}M_\odot$
\citep{pach07} --- so if the companion is larger than a typical
low-mass white dwarf then it is likely extremely young, and we might
be seeing a newly-born millisecond pulsar slightly after the
transitional phase found by \citet{asr+09}.  The hot/large state could
also be a result of a recent hydrogen flash.  This source then
resembles the rather young WD/sdB stars \object[HD 188112]{HD~188112}
($T_{\rm eff}=$21,500\,K, $M=0.24\,M_\odot$, $R=0.1\,R_\odot$;
\citealt{heln03}) or \object[GALEX J1717+6757]{GALEX~J1717+6757}
($T_{\rm eff}=$14,900\,K, $M=0.19\,M_\odot$, $R=0.1\,R_\odot$;
\citealt{vtk+11}) which are thought to be progenitors of more typical
helium-core white dwarfs.  Both of these scenarios (mass stripping or
hydrogen flash) are possibilities for the hot, bloated white dwarfs
seen by \textit{Kepler} (\citealt{vkrb+10}; \citealt*{crf11};
\citealt{brvkc12}), which the companion to \psr\ also resembles,
although the presence of a neutron star instead of a main sequence
primary would require a different evolution.  Again, spectroscopy to
determine surface gravity and elemental abundances should be
definitive.  Finally, it is possible that, despite the eclipses the
system is closer to face-on than edge on, as for inclinations
$<30\degr$ the companion mass is $>0.5 M_\odot$ like other hot white
dwarf companions.  This would be contrary to the theoretical companion
mass vs.\ orbital period relation \citep{ts99} and would make the
radius even stranger, but there is at least one known outlier from the
companion mass vs.\ orbital period relation with a substantially
higher mass than expected \citep{hrr+05,dpr+10}.

The bright counterpart makes optical astrometry within the reach of
ground-based telescopes, at least for determining a proper motion.
While we see no definitive proper motion comparing the \swift\ and
WIYN data (taken 0.6\,yr apart), individual ground-based images can
determine the relative position of the pulsar to $\lesssim 10\,$mas in only a
few minutes.  We expect a proper motion $\mu=10 v_{100} d_2^{-1}\,{\rm
  mas\,yr}^{-1}$, so assuming adequate calibration this can be
measured in a year or two.  We could then compare this against any
radio proper motion determined from timing, which would further
establish whether or not the optical source is indeed the companion of
the pulsar.

\section{Conclusions}
\label{sec:conc}
We have discovered the almost certain optical/UV counterpart of the
newly-discovered millisecond radio/$\gamma$-ray pulsar \psr.  The
radio/$\gamma$-ray properties of \psr\ appear much like most energetic
eclipsing pulsars discovered recently \citep{roberts11}, which
supports the use of \fermi\ and low-frequency radio observations to
find energetic recycled pulsars.  Some aspects of \psr's radio
properties appear unique, in that the plausible size of the eclipsing
region seems to be contained in its companion's Roche lobe, but such
inferences depend on the inclination as well as the observing
frequency of the radio data.

While we still cannot constrain all of its parameters uniquely, the
optical/UV properties of \psr\ appear more robustly unique, with an
effective temperature likely at least 3 times higher than any other
black-widow or redback system.  In fact, the companion may be the
brightest low-mass (i.e., not a B star) optical\footnote{It is
  expected to be as bright as the companion to PSR~J0437$-$4715 in the
  near-infrared \citep{dkp+11}, consistent with its non-detection in
  2MASS.}  companion to any pulsar \citep{vkbjj05}.  This is largely
because of the high temperature rather than a small distance or large
size.  The high temperature presents a number of puzzles and
opportunities.  Depending on the radius, it may be that the companion
is rather young, and that we are seeing the sources only shortly after
its envelope was stripped away.  Phase-resolved photometry and
spectroscopy will be important to determine the orientation and mass
function of the system \citep{vkbk11}, and radio astrometry can help
constrain the radius of the companion. Long-term optical monitoring
may be able to detect cooling after a recent stripping or burning
episode. Given how bright it is, modulation of the companion may be
detectable from a number of mechanisms.  Orbital motion may be visible
through Doppler boosting \citep{vkrb+10,sks+10b}, which is expected to
produce modulation of $\pm0.3$\%, while ellipsoidal modulations ($\pm
0.1$\%) could help constrain the mass ratio and/or the degree of
Roche-lobe filling (\citealt{vkrb+10}; \citealt{crf11};
\citealt{brvkc12}).  At the same time, the high temperature and small
size compared to the Roche lobe mean that the photocenter may be much
closer to the geometric center of the star \citep[cf.][]{vkbk11}
facilitating (along with its brightness) measurement and
interpretation of the radial velocity curve, and with it enabling an
accurate mass measurement for this neutron star.

\acknowledgements We thank an anonymous referee for useful comments,
P.~Ray, L.~Bildsten, S.~Phinney, R.~Breton, M.~van~Kerkwijk, and
C.~Bassa for helpful discussions, and D.~Koester for supplying some
white dwarf atmosphere models.  Based upon data from the WIYN
Observatory, which is a joint facility of the University of
Wisconsin-Madison, Indiana University, Yale University and the
National Optical Astronomy Observatories.  The National Radio
Astronomy Observatory is a facility of the National Science Foundation
operated under co-operative agreement by Associated Universities, Inc.
Some of the data presented in this paper were obtained from the
Multimission Archive at the Space Telescope Science Institute
(MAST). STScI is operated by the Association of Universities for
Research in Astronomy, Inc., under NASA contract NAS5-26555. Support
for MAST for non-HST data is provided by the NASA Office of Space
Science via grant NNX09AF08G and by other grants and contracts. JWTH
is a Veni Fellow of the Netherlands Foundation for Scientific Research
(NWO).  CMB, DFD, MDWR, and XS were supported by NSF CAREER award
number 09955929 and PIRE award number 0968126, with additional support
from the University of Wisconsin--Milwaukee Office of Undergraduate
Research. Pulsar research at UBC is supported by an NSERC Discovery
Grant.

{\it Facilities:} \facility{Fermi (LAT)}, \facility{Swift (UVOT, XRT)},
\facility{GALEX}, \facility{WIYN (MiniMo)}

%\bibliography{psr}

%% --------------------------------------------------------------------
%% Mon Apr  9 10:35:14 2012
%%   This file was generated automagically from the files
%%   psrj1816.bbl and psrj1816.tex using
%%     /Users/dlk//perl/nat2jour.pl
%%   This file should accompany psrj1816-aas.tex.
%% --------------------------------------------------------------------

\begin{deluxetable}{l c c c c c c c c}
\tablewidth{0pt}
\tabletypesize{\footnotesize}
\setlength{\tabcolsep}{2pt}
\tablecaption{Photometry of the Optical Counterpart to \psr\label{tab:phot}}
\tablehead{
\colhead{Band\dotfill} & \colhead{FUV} & \colhead{NUV} & \colhead{$u$} & \colhead{$B$} &
\colhead{$V$} & 
  \colhead{$R$} & \colhead{$R$} & \colhead{$I$} 
}
\startdata
Magnitude\dotfill & $19.86\pm0.18$ & $19.11\pm0.09$ & $18.52\pm0.04$ &
$18.3\pm0.2$ &
$18.51\pm0.06$ & $18.4\pm0.2$ & $18.27\pm0.04$ & $18.4\pm0.2$\\
Wavelength (nm)\dotfill & 152 & 230 & 345 & 438 & 545 & 641 &
641 & 798\\
Source\dotfill & \galex & \galex & \swift & DSS & CSDR1 & DSS & WIYN & DSS\\
System\dotfill & AB & AB & AB & Vega & Vega\tablenotemark{a} & Vega & Vega & Vega\\
\enddata
\tablecomments{For each measurement, we give the measured magnitude
  with uncertainties (assumed to be 0.2\,mag for the DSS), the
  approximate central wavelength, the origin (telescope/survey), and
  whether the magnitude is on the AB or Vega system.}
\tablenotetext{a}{This was transformed from the unfiltered detector
  system assuming $B-V=0$.}
\end{deluxetable}

\end{document}